# Voice Conversion with Conditional SampleRNN

*Cong Zhou, Michael Horgan, Vivek Kumar, Cristina Vasco, Dan Darcy*

Dolby Laboratories, San Francisco, CA, USA
{cong.zhou, mhorg, vivek.kumar, cvasc, dan.darcy}@dolby.com

## Abstract

Here we present a novel approach to conditioning the SampleRNN [1] generative model for voice conversion (VC). Conventional methods for VC modify the perceived speaker identity by converting between source and target acoustic features. Our approach focuses on preserving voice content and depends on the generative network to learn voice style. We first train a multi-speaker SampleRNN model conditioned on linguistic features, pitch contour, and speaker identity using a multi-speaker speech corpus. Voice-converted speech is generated using linguistic features and pitch contour extracted from the source speaker, and the target speaker identity. We demonstrate that our system is capable of many-to-many voice conversion without requiring parallel data, enabling broad applications. Subjective evaluation demonstrates that our approach outperforms conventional VC methods.

## 1. Introduction

Voice conversion (VC) commonly refers to the task of transforming a speech signal such that the identity of the source speaker is changed to that of a target speaker while all linguistic content remains unchanged. While one could argue that the linguistic content, or meaning, of speech can be captured by a combination of phonemes, inflection, and cadence, the concept of voice identity is harder to codify. Nonetheless, traditional VC approaches have focused on modeling the precise sequence of both linguistic and identity-specific (i.e. voice-stylistic) events in the form of hand-crafted frame-based acoustic features. Not surprisingly, while linguistic content generally remains accessible in the target speech, traditional VC systems struggle to accurately mimic the target speaker identity or preserve naturalness [2]. We propose an alternative approach to modeling speaker identity via deep autoregressive models. The approach is both simpler to implement and more effective, eliminating the need to manually codify speaker identity via hand-engineered features while also improving naturalness and perception of speaker identity.

A typical VC system relies on conversion techniques, such as Gaussian Mixture Models (GMMs), to learn a mapping of vocoder features (e.g. spectral envelope, formants, mel-cepstrum, etc.) from a source speaker to a target speaker [2, 3]. These systems frequently require training on parallel datasets, where each utterance is spoken by both source and target speakers. During training, features are extracted from both signals and time-aligned to compensate for differences in the cadence of the linguistic events. Then the model is optimized (e.g. for maximum likelihood) to learn a mapping from source to target features that ideally preserves the linguistic content that is common to both feature sets while converting the voice stylistic information into a form that is likely to have been extracted from the target signal. Within this workflow, there are several points that are prone to error and which may help to explain the deficiencies observed in traditional VC, including: whether the chosen acoustic features capture all of the information that is relevant to voice style and/or linguistic content; the accuracy of the time-alignment algorithm; the assumption that time-warping has no impact on the distribution of voice style features; and for models that require parallel datasets, the implicit assumption that for each source feature, there is exactly one ground-truth target feature (when in reality the target speaker is likely capable of producing a broad distribution of stylistic features for the same linguistic content).

Recently deep neural networks have been utilized for VC. For example, [4] employs the popular WaveNet [5] neural audio generative model as the vocoder in a VC system, conditioning it on acoustic features that have been converted from source to target using a GMM. [6] also introduces a novel VC method that utilizes an autoencoder architecture for separating content and style into separate latent variables. Once separated, the style variables can be swapped with those of a different speaker and the resulting hybrid representation can then be decoded to achieve voice conversion. However, the audio quality reported by these systems remains substantially lower than natural human speech. Therefore, it is worth investigating methods to further improve speech quality.

SampleRNN is a deep neural audio generative model recently proposed in [1] for generating high quality audio without conditioning. Conditional variants have also been developed; for example, [7] introduces a conditional SampleRNN that serves as the vocoder within a text-to-speech system. Drawing inspiration from [7], we develop a text-dependent conditional SampleRNN and use it as the target speaker generator within a VC system. But rather than predicting alignment and pitch information from text, we infer these parameters directly from the source utterance. In contrast to traditional VC and to [4], we do not condition the model on hand-crafted acoustic features meant to capture stylistic information (with the exception of F0, which arguably contains both linguistic and stylistic information). Rather, the model learns to extract stylistic information on its own through the optimization process. Furthermore, there is no requirement for source-target parallel datasets or forced alignment, as the training process focuses solely on learning to mimic a target speaker (or set of target speakers), rather than learning to convert one speaker's style to another's. The subjective evaluation results described in Section 4 show that our system outperforms the conventional VC baseline.

Similar to [6], we also draw inspiration from [8], and although the final form of our solution differs significantly, we credit [8] with inspiring us to use a deep neural network as the principal tool for decoupling content from voice style.

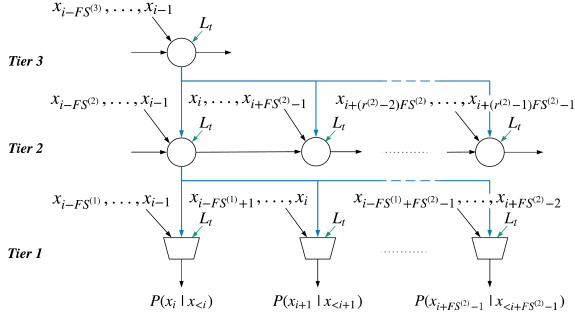

Figure 1: *Unrolled structure of conditional SampleRNN at timestamp i. Blue arrows represent conditioning information from upper tier. Green arrows with $L_t$ represent linguistic conditioning context at frame t, where $t = \lfloor i/FS^{(3)} \rfloor$.*

## 2. Conditional SampleRNN

### 2.1. SampleRNN

SampleRNN [1] introduced an architecture for unconditional neural audio generation that is capable of modeling the joint distribution of high-dimensional audio data (e.g. 8-bit 16 kHz) via factorization of the joint distribution into the product of the individual audio sample distributions conditioned on all previous samples. The joint probability of a sequence of waveform samples $X = \{x_1, ..., x_T\}$ can be written as

$$p(X) = \prod_{i=0}^{T-1} p(x_{i+1}|x_1, ..., x_i) \quad (1)$$

The architecture consists of a series of multi-rate recurrent layers followed by a series of fully-connected layers. The layers are grouped into tiers according to the rate at which they operate, where the lowest tier consists of the fully-connected layers and operates at the maximal rate of one audio sample. Each higher tier consists of one or more RNN layers and operates at a progressively lower rate. The highest tier runs first and conditions the input to the next tier with its output. This process is repeated for all tiers in the network. The entire model is trained jointly using truncated back-propagation through time (TBPTT) to minimize the negative log likelihood between the predicted distribution and the target distribution.

### 2.2. Conditional SampleRNN

When trained solely on audio samples of human speech, SampleRNN learns to generate realistic-sounding "babble", or unintelligible vocal utterances. In this paper, we present a method for conditioning SampleRNN with high-level linguistic information that enables the generation of intelligible speech. Eq. (1) becomes

$$p(X) = \prod_{i=0}^{T-1} p(x_{i+1}|x_1, ..., x_i, \boldsymbol{h}) \quad (2)$$

where $\boldsymbol{h}$ represents linguistic features. The structure of conditional SampleRNN is illustrated in Figure 1. Similar to SampleRNN, the $k$-th tier $(1 < k \leq K)$ of conditional SampleRNN operates on non-overlapping frames of size $FS^{(k)}$ samples at a time, and the lowest tier $(k = 1)$ module operates at sample-level. The upper tiers are implemented with RNNs, and the lowest tier is a multilayer perceptron (MLP). $r^{(k)}$ is the ratio between the temporal resolution of tier $k + 1$ and tier $k$.

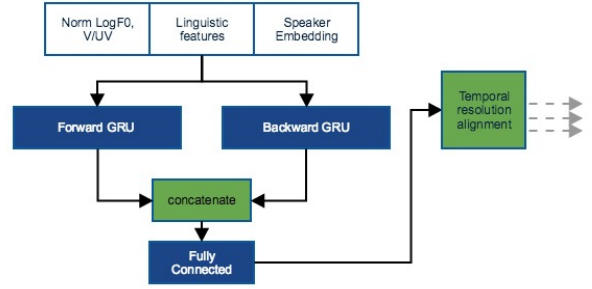

Figure 2: *High level structure of conditioning network*

$L$ is a sequence of linguistic conditioning context from the conditioning network, and is added to the input of tier $k$ as auxiliary input features. The $k$-th tier $(1 \leq k < K)$ has three inputs: audio samples, conditioning information from tier $k + 1$, and conditioning context from the conditioning network. The top tier has two inputs: audio samples and conditioning context from the conditioning network.

Inspired by the work of Deep Voice [9], where a stack of bidirectional quasi-recurrent networks for conditioning Wavenet was introduced, we proposed one single layer bidirectional Gated Recurrent Unit (GRU) [10] with a linear layer as our conditioning network. Similar to Deep Voice 2 [11], the input to the conditioning network contains linguistic features, pitch contour and speaker embedding. A diagram of the conditioning network is shown in Figure 2. The temporal resolution alignment step is required because the output of the conditioning network may have a different temporal resolution from the tiers in conditional SampleRNN. Upsampling by repetition is used in our system, and a learned upsampling method could be explored further.

#### 2.2.1. Single-speaker model

The two main conditioning data that we use are phoneme labels and fundamental frequency (F0) contour. Prior to training the model, the dataset, which consists of single-speaker speech signals on the order of 2 to 8 seconds in length and the corresponding text transcriptions of those sentences, is preprocessed using existing tools [12, 13] to extract the phonemes, with alignment times, and frame-based log-scaled F0 of each signal. We then construct a conditioning signal at a 200 Hz sample rate, wherein each vector is a concatenation of: 1) one-hot vectors of the previous two, current, and next two phonemes (i.e. 5 total one-hot vectors), 2) the scalar log F0 value, 3) a scalar Boolean indicating whether the corresponding current phoneme is voiced or unvoiced (i.e. whether the F0 is relevant to the synthesis of the current phoneme).

At training time, we reset the hidden state of the backward portion of the bidirectional GRU once per gradient step; while at generation time, we only reset it once per utterance (as we do for all other hidden states in the model).

#### 2.2.2. Multi-speaker model

We also demonstrate that the single-speaker conditional SampleRNN model can be extended to a multi-speaker model by further conditioning each tier on a speaker ID, similar to [11]. We use a 16-dimensional learned embedding table to represent the speaker IDs and concatenate the speaker embedding to each sample in the conditioning signal prior to processing it with the bidirectional GRU.

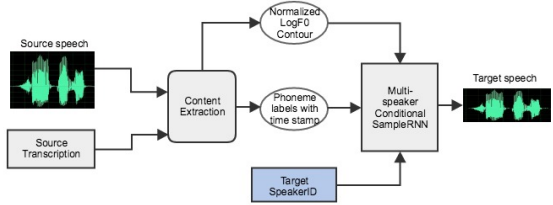

Figure 3: *Overview of conditional SampleRNN based voice conversion system*

# 3. Voice style transfer with conditional SampleRNN

In this section, we propose the system to generate voice converted waveform samples based on conditional SampleRNN described in Section 2.2. Figure 3 illustrates the conversion process using a trained multi-speaker conditional SampleRNN model.

## 3.1. Content extraction

Prior to the inference procedure, aligned phoneme labels and log-scaled F0 values are extracted from the source speech signal using existing tools [12, 13]. This information represents the content from the source signal that is preserved in the voice-converted target signal.

Frame-by-frame linear transformation of log-scaled F0 is a common technique [3, 14] used to resolve pitch differences between the source speaker and target speaker. We adopted this technique and normalized each speaker's log-scaled f0 frame-by-frame to have zero mean and unit variance:

$$x'_t = \frac{x_t - \mu^{(x)}}{\sigma^{(x)}} \quad (3)$$

where $x_t$ is the log-scaled F0 of a particular speaker at frame $t$, $\mu^{(x)}$ and $\sigma^{(x)}$ are the mean and standard deviation of log-scaled F0 of this particular speaker, and $x'_t$ is normalized log-scaled F0.

We used Eq. (3) in both training and inference (VC) stage. In the training stage, $x'_t$ represents the normalized log-scaled F0 from the target; in the inference stage, $x'_t$ represents the normalized log-scaled F0 from the source. The model learns to associate a given speaker's embedding vector with a characteristic mean and variance for that speaker's log-scaled F0 distribution, thus allowing the model to translate arbitrary, normalized log-scaled F0 contours into a vocal range that is appropriate for the target speaker.

## 3.2. VC via target speaker generation

Once we have a fully trained SampleRNN model that has been conditioned on aligned phonemes, normalized log-scaled F0 contours, and speaker IDs, we can utilize the model to generate voice-converted speech for any speaker in the training set. When the model is conditioned on the aligned phoneme labels and normalized log-scaled F0 contours of the source speaker's utterance, along with the speaker embedding associated with the desired target speaker, it generates a vocal utterance that has the same linguistic content, alignment, and normalized log-scaled F0 contour as that of the source speaker's utterance with the target speaker's voice. The conditioning inputs are easily extracted from any speaker, so the system does not require the source speaker to be in the training set. Since the system is based on the target speaker generator, rather than on the mapping between source and target speakers, it doesn't rely on a source-target parallel dataset.

# 4. Experimental evaluation

## 4.1. Experimental setup

We evaluated converted speech naturalness and speaker similarity to compare the performance of the conventional method [3] and proposed method. We used [15] as the implementation of the conventional method. Although our system accepts non-parallel speech data, we used fully parallel data for comparing between two methods. The CMU-Arctic dataset [16] was used, and "bdl" and "clb" were chosen as source and "rms" and "slt" were chosen as target. We defined four pairs of conversions, two intra-gender conversions and two cross-gender conversions, and used 1028 sentences for training and 104 sentences for evaluation. Training files were converted into 0-255 discrete values by mu-law quantization. All the speech files have 16 kHz sampling rate.

In the feature extraction stage, WORLD [13] was used to extract pitch, and the frame size was 5 ms; phoneme labels with alignments were included within the Arctic dataset.

In the training stage, the multi-speaker conditional SampleRNN model described in section 2.2.2 was jointly trained for all 4 speakers. SampleRNN was configured into three tiers with different temporal resolutions in each tier. The top tier RNN had frame size 5 ms, 80 samples; the middle tier RNN had frame size 0.5 ms, 8 samples; the bottom MLP layer had 8 samples look-ahead and long-term contexts from upper tiers. We used 1 GRU layer for top tier and middle tier RNNs, with 1024 hidden units. We constructed the bottom MLPs the same as the original SampleRNN, with 3 fully connected layers with ReLU activation, an output dimension of 1024 for first two layers and 256 for the final layer before a final 256-way softmax layer. We used 1 GRU layer for the forward RNN and 1 GRU layer for the backward RNN in the conditioning network, with the hidden units of both forward GRU and backward GRU set to 1024. The conditioning bidirectional GRU operated on 5ms temporal resolution, and upsampling through repetition was used to match the temporal resolution between the bidirectional GRU and each tier of the SampleRNN. Negative Log-Likelihood in bits per audio sample was used for a loss function. Adam optimizer [17] with a constant learning rate of 0.001 was used to adjust the parameters. Mini batch size was 32 clips, each clip has 4 seconds, and 8000-sample sequence lengths were used for TBPTT.

## 4.2. Subjective evaluation

To assess the performance of our model across a subset of content from the CMU-Arctic dataset for the four intra- and cross-gender conversions, we conducted human subjective assessment experiments. Our test design was adapted from ITU-R BS.1116-3, a standard ratings methodology for measuring small impairments in audio systems based on how impairment artefacts range from imperceptible, to perceptible but not annoying, to slightly and finally highly annoying.

We tested the systems' ability to produce natural-sounding speech by altering the original BS.1116-3 design, making explicit reference to the naturalness of a system under test rather than annoyance due to degradation. Listeners were presented with a reference system (the target) and two systems under test:

a hidden reference, and either our conditional SampleRNN model or a mixture-of-Gaussians model here called "Baseline" [10]. Trials were randomized and system identification was masked for both subjects and test administrators.

Nine listeners (three female) participated in the speech system evaluation. Listeners were instructed to identify the hidden reference and score it the maximum rating of 5. These data demonstrate whether the conditional SampleRNN and Baseline systems are reliably differentiated from reference. Next, listeners rated the other system using the following scores and verbal descriptor (Table 1).

Table 1: *Verbal descriptor and rating scores for systems under test.*

| Verbal Descriptor | Rating Score |
|---|---|
| Imperceptible | 5 |
| Perceptible, but not unnatural | 4 |
| Slightly unnatural | 3 |
| Unnatural | 2 |
| Highly unnatural | 1 |

These data were used to determine the relative performance of conditional SampleRNN and Baseline systems based on their ability to produce natural-sounding speech.

We also designed a forced-choice test to assess similarity of speech produced by a system relative to a reference target. For this experiment, users were presented a reference target and a paired comparison, ours and the Baseline system, and instructed to choose which system produced speech most similar to reference. These results are described below.

### 4.3. Results

Our data evaluating naturalness of systems indicate that our conditional SampleRNN system consistently outperforms Baseline across the selected 9 clips from the CMU-Arctic data set for all four intra- and cross-gender conversions. Because the hidden reference was always correctly identified and scored a 5, indicating that for no trials was either our system or Baseline audio indistinguishable from the unprocessed reference target for any listener, only system data are presented for clarity. Figure 4 shows data for the four conversion conditions across the 9 CMU-Arctic clips.

Additionally, we conducted a forced-choice assessment in which subjects were asked to choose which system, conditional SampleRNN or Baseline, was most similar in voice identity to the unprocessed target. After preliminary data collection, we observed in 100% of trials that conditional SampleRNN was judged more similar to target voice identity than Baseline (data not shown), so we accepted these results as evidence that future studies in this area will require system comparisons that are more closely matched in performance than the systems under test in the present study.

Together, these results demonstrate that our conditional SampleRNN system outperforms the Baseline system in naturalness of speech. All statistically significant results were a result of higher scores for our system than Baseline. Averaging across content (Table 2), all conversion conditions were rated very close to or higher than a 3 for conditional SampleRNN, indicating systems ranged from 'slightly unnatural' towards an upper-bound near 4 which is the crossing point for 'perceptible, but not unnatural'. In contrast, the Baseline system scored very close to or lower than a 3 which crosses into the 'unnatural' descriptor.

Table 2: *Mean results across content of SampleRNN and Baseline speech synthesis transfer systems*

| Conversion | SampleRNN | Baseline |
|---|---|---|
| F1-to-F2 | 3.73±0.14 | 3.04±0.18 |
| M1-to-M2 | 3.01±0.18 | 2.49±0.18 |
| F1-to-M2 | 2.95±0.19 | 2.29±0.14 |
| M1-to-F2 | 3.19±0.20 | 2.35±0.19 |

Interestingly, each system achieved its highest score by a substantial margin for the female1 to female2 conversion. Other systems tended to score more closely together. This suggests that cross-gender conversion is not an innately more challenging condition for either system, or that there are uniquely relevant features used by the systems when a male voice or female voice are targeted.

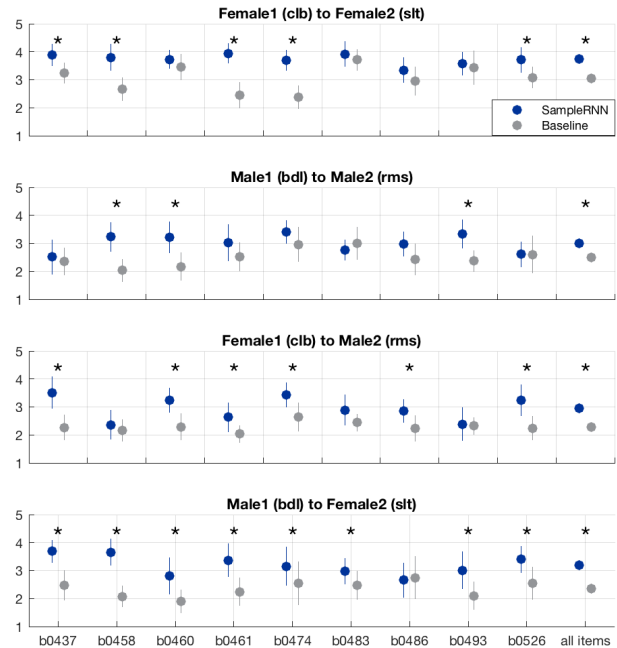

Figure 4: *Results for 9 CMU-Arctic dataset clips used in four intra- and cross-gender conversion conditions. Mean results for all items are in the rightmost column. Asterisks denote statistical significance (p<0.05) using Student's t-test.*

## 5. Conclusions

In this paper, we proposed a text-dependent VC method based on the conditional SampleRNN speech generative model. The advantage of the proposed method is that it does not require a parallel dataset and it does not have restrictions on source speakers. The experimental results demonstrate that the proposed method has a significant improvement on speech naturalness over the baseline system with much higher conversion accuracy on speaker identity. In future work, we plan to further improve speech quality and develop an end-to-end text-independent voice conversion method.

## 6. Acknowledgements

The authors thank Grant Davidson, Lie Lu, and Nicolas Tsingos for their helpful feedback, Rich Graff and Nathan Swedlow for data collection and administration, as well as Jaime Morales for early discussions.